\documentstyle[aps,floats,amssymb,12pt]{revtex}

\input BoxedEPS
\SetRokickiEPSFSpecial  %% for dvips by Tom Rokicki, for VMS
\HideDisplacementBoxes

\begin{document}

\newcommand{\pval}{{-}\mkern-18.5mu\int}
\newcommand{\pvalt}{\raise0.15ex\hbox{-}\mkern-12mu\int}

\newcommand{\be}{\begin{equation}}
\newcommand{\ee}{\end{equation}}
\newcommand{\bea}{\begin{eqnarray}}
\newcommand{\eea}{\end{eqnarray}}
\newcommand{\ben}{\begin{enumerate}}
\newcommand{\een}{\end{enumerate}}
\newcommand{\bit}{\begin{itemize}}
\newcommand{\eit}{\end{itemize}}
\newcommand{\la}[1]{\label{#1}}
\newcommand{\dk}{\frac{d^{4}k}{(2\pi)^4i}}
\newcommand{\di}[1]{\frac{d^{4}{#1}}{(2\pi)^4i}}
\newcommand{\eq}[1]{eq.~(\ref{#1})}
\newcommand{\Eq}[1]{Eq.~(\ref{#1})}
\newcommand{\eqs}[2]{eqs.~(\ref{#1}) and (\ref{#2})}
\newcommand{\Eqs}[2]{Eqs.~(\ref{#1}) and (\ref{#2})}
\newcommand{\f}[2]{\tilde {F}_{#1}(#2)}
\newcommand{\half}{{\textstyle\frac{1}{2}}}
\newcommand\svec[1]{\skew{-2}\vec#1}
\newcommand{\ns}{\hspace{-0.5ex}}
\newcommand{\p}{\partial}
\newcommand{\hf}{{\textstyle \frac{1}{2}}}
\newcommand{\Tr}{\mathop{\mathrm Tr}}
\newcommand{\trc}{\,{\mbox{Tr}}_C\,}
\newcommand{\tr}{\,\mbox{tr}\,}
\newcommand{\Ln}{\,\mbox{Ln}\,}
\renewcommand{\ln}{\,\mbox{ln}\,}
\renewcommand\Im{\mathop{\mathrm Im}}
\newcommand\mn{{\mu\nu}}
\newcommand\parm{\par\medskip}
\renewcommand\Re{\mathop{\mathrm Re}}

\newcommand{\cl}{\centerline}
\renewcommand{\thefootnote}{\fnsymbol{footnote}}

\renewcommand{\baselinestretch}{1.0}

\title{The Casimir Energy in a Separable Potential\thanks{This work is
supported in part by funds provided by the U.S. Department of Energy
(D.O.E.) under cooperative research agreement \#DF-FC02-94ER40818.}}

\author{R.L.~Jaffe\thanks{email: jaffe@mit.edu} and
L.R.~Williamson\\[1ex] }

\address{Center for Theoretical Physics\\
Laboratory for Nuclear Science\\
and Department of Physics \\
Massachusetts Institute of Technology\\
Cambridge, Massachusetts 02139\\[1ex]
{\rm (MIT-CTP Preprint 2881, hep-th/9907199\quad July 1999)\\
Submitted to \emph{Annals of Physics} }}

\maketitle

\bigskip

\begin{abstract}
The Casimir energy is the first-order-in-$\hbar$ correction to the energy
of a time-independent field configuration in a quantum field theory.  We
study the Casimir energy in a toy model, where the classical field is
replaced by a separable potential.  In this model the exact answer is
trivial to compute, making it a good place to examine subtleties of the
problem.  We construct two traditional representations of the Casimir
energy, one from the Greens function, the other from the phase shifts,
and apply them to this case.  We show that the two representations are
correct and equivalent in this model.  We study the convergence of the
Born approximation to the Casimir energy and relate our findings to
computational issues that arise in more realistic models.
\end{abstract}


\pagestyle{myheadings}
\markboth{The Casimir Energy in a Separable Potential}{R.L.~Jaffe and
L.R.~Williamson}
\thispagestyle{empty}

\setlength{\baselineskip}{0.75\baselineskip}
\vspace*{-3pc}

\section{Introduction}\label{Introduction}
Casimir energies are important in quantum field theories, where they 
give the first-order (in $\hbar$) corrections to the energy of 
time-independent classical field configurations.  Computations of the 
Casimir energy involve formal manipulations of divergent expressions 
that eventually are regulated and renormalized.\cite{fghj} Since they 
go beyond perturbation theory, Casimir energy calculations provide a 
potentially powerful way to study nonperturbative effects.  Several 
formally equivalent representations of the Casimir energy have been 
used in numerical calculations, notably (a)~in terms of integrals over 
scattering phase shifts and bound states; (b)~as an integral over the 
trace of a Greens function (equivalent to a functional determinant); 
and (c)~as an infinite sum of Feynman diagrams.  The aim of this paper 
is to study the various representations of the Casimir energy in a 
simple, highly convergent toy model, where the equivalence of 
different representations can be demonstrated without the complication 
of divergences.  Our analysis does not demonstrate the equivalence of 
representations in realistic models, where divergences make the 
arguments more complex.  However, we do see how the different 
representations are related in a simple, calculable example.  Also, 
because the model is so simple, it is possible to explore some issues 
of convergence and cancellation that are obscure in more realistic 
cases.

The model we study makes use of the simple dynamics of separable 
potentials in nonrelativistic quantum mechanics.\cite{Gottfried} Our 
first task is to explain how such a simple model can give insight into 
one-loop effects in a quantum field theory.  Next we review the 
well-known Greens function and scattering theory analysis of a 
separable potential.  With this in hand we discuss the various 
representations of the Casimir energy and explore their relationship.  
Finally, we study convergence of the perturbative expansion and 
associated computational issues.

Formally, the Casimir energy is the sum of zero point energies for the 
modes of a quantum field, $\Psi$, in the presence of some spatially 
varying background, $\phi_{0}(x)$,
\begin{equation}
    {\cal E}_{c} = \half\sum(\hbar\omega[\phi_{0}]-\hbar\omega_{0})
    \la{0.1}
\end{equation}
where $\{\omega[\phi_{0}]\}$ are the eigenfrequencies in the presence 
of the nontrivial background field $\phi_{0}$, and $\{\omega_{0}\}$ 
are the eigenfrequencies in the trivial background, $\phi_{0}=0$.  The 
$\{\omega[\phi_{0}]\}$ are typically related to the eigenvalues of 
some simple differential operator, ${\cal H}[\phi_{0}]$, which looks 
like a Schr\"odinger operator for a scalar field,
\begin{equation}
    {\cal H} = -\frac{d^{2}}{dx^{2}} + V_{\phi_{0}}(x) 
    \la{0.2}
\end{equation}
or a Dirac operator for a spinor field.  $V_{\phi_{0}}(x)$ is a 
``potential'' derived from the field $\phi_{0}$.  Although we have 
written eq.~(\ref{0.2}) for a scalar in one-dimension, the same 
considerations apply in three-dimensions.  Typically the $\{\omega\}$ 
are not themselves the eigenvalues, $\{\lambda\}$, of ${\cal H}$ but 
rather simple functions of the them.  For example, in the case that 
$\Psi$ is a scalar, $\omega_{j}=\sqrt{\lambda_{j} + m^{2}}$.

Our toy model is based on two alterations in this physical picture: 
first we assume that the $\{\omega\}$ are proportional to the 
eigenvalues of ${\cal H}$, as they would be in the nonrelativistic 
case where $\hbar\omega(k)= \hbar^{2}k^{2}/2m$ (and $k^{2}$ is the 
eigenvalue of ${\cal H}$); and second, we replace the \emph{local} 
potential, $V(x)$, by a nonlocal, but \emph{separable} potential,
\be
        V(x,x')=-\lambda u(x)u(x')
        \la{0.21}
\ee
where $\int dx \,  u^{2}(x) = 1$.  Because of our first assumption the 
Casimir energy can be written as
\be
    {\cal E}_{c} = \half \Tr ({\cal H} - {\cal H}_{0})\ .
    \la{0.3}
\ee
As a consequence of the separability assumption, ${\cal H}$ can be
written formally as
\be
     {\cal H} = {\cal H}_{0} - \lambda |u\rangle\langle u| 
     \la{0.4}
\ee
where $|u\rangle$ is the state with wavefunction $u(x)=\langle 
x|u\rangle$.  Combining eqs.~(\ref{0.3}) and (\ref{0.4}) we find that 
the Casimir energy is $-\half\lambda$,
\bea
        {\cal E}_{c} &=& -\half \Tr \{ -\lambda |u\rangle\ \langle u|\}
        \nonumber\\
        &=& -\half\lambda\ .
        \la{0.5}
\eea
This is the fundamental result that makes the study of the Casimir 
energy in separable potential models interesting -- the answer is so 
transparent.  Our object in this paper is to show that more 
conventional (and more complicated) methods of computing the Casimir 
energy coincide with this simple result.

The remainder of the paper is organized as follows.  In the next 
section we review the bound states, Greens functions and scattering 
amplitude for a separable potential.  In Section~\ref{CasimirEnergy} we
derive  expressions for the Casimir energy in terms of integrals over
Greens  functions or scattering phase shifts.  In
Section~\ref{CasimirEnergySeparable} we compute the  Casimir energy
using the methods of Section~\ref{CasimirEnergy}  for the case of a 
separable potential and show that the result is $-\half \lambda$.  In 
Section~\ref{Convergence} we specialize to a particular choice of $u$
and explore the  convergence of the Casimir energy calculation as a
function of the  strength, $\lambda$, of the interaction.  We conclude in
Section~\ref{Summary}.

%%%%%%%%%%%%%%%%%%%%%%%%%%%%%%%%%%%%%%%%%%%%%%%%%%%%%%%%%%%
\section{Separable Potentials}\label{Separable}

In this section we review the solution of the scattering problem for a
scalar particle moving in an s-wave separable potential in three
dimensions.  We begin with the Schr\"odinger equation in three
dimensions with a nonlocal potential,
\be
   -\frac{\hbar^2}{2M}\nabla^2\psi(\svec r) +\int d^3 r^{\prime}\,
V(\svec 
   r,\svec r^{\prime} ) \psi(\svec r^{\prime} ) = \omega \psi(\svec r ) 
   \la{1.3.1}
\ee
where
\be
  \langle\svec r|V|\svec r^{\prime} \rangle = V(\svec r,\svec
  r^{\prime} )\ .
  \la{1.3}
\ee
The resulting integral equation is in general more complicated than 
the local case, but for a separable potential, \eq{0.21}, it is 
simpler.  For simplicity we take a \emph{separable and spherically 
symmetric} nonlocal potential,
\be
	V(\svec r,\svec r^{\prime} ) = -\lambda u(r)u(r^{\prime} )
	\la{1.4}
\ee
for which \eq{1.3.1} reduces to
\be
	-\frac{\hbar^2}{2M}\nabla^2\psi(\svec r) -\lambda u(r)\int
	d^{3}r^{\prime}\, u(r^{\prime})
	\psi(\svec r^{\prime} ) =  \omega  \psi(\svec r )\ .
	\la{1.3.2}
\ee
\nopagebreak
For $\lambda>0$ the potential is attractive.

A principal simplification with a spherically symmetric, separable 
potential is the absence of any interaction in partial waves with 
$\ell>0$.  This follows immediately from \eq{1.3.2} because $\int 
d^3r^{\prime}\, u(r^{\prime} )\psi(\svec r^{\prime} )$ projects out only 
the spherically symmetric part of $\psi$.  All partial waves except 
$\ell=0$ cancel out of the Casimir sum.  We replace $\psi(\svec r)$ by
$U(r)/r$ and find that \eq{1.3.2} simplifies to
\be
  	-U''(r) -4\pi\lambda ru(r)\int_0^\infty\!\! dr^{\prime}
	r^{\prime}  u(r^{\prime} ) U(r^{\prime} ) = k^2 U(r)
  	\la{2.1}
\ee
where $k^2 = 2M\omega/\hbar^2$, and where we have set $\hbar=2M=1$ 
henceforth.

For scattering solutions, it is useful to convert \eq{2.1} into an 
integral equation for the function, $U^{(+)}(r)$, obeying scattering 
boundary conditions,
\be
	U^{(+)}(r) = U_0(r) +4\pi\lambda\int_0^\infty \!\!
	dr' G_0^{(+)}(r,r',k)r' 
  	u(r') \int_0^\infty dr''r''u(r'')U^{(+)}(r'')
  	\la{2.4}
\ee
where $U_0(r)$ is the free solution regular at the origin, $U_0(r) = 
\sin kr$, and $G_0^{(+)}(r,r^{\prime},k)$ is the free, s-wave Greens 
function with outgoing wave boundary conditions,
\bea
  	G_0^{(+)}(r,r^{\prime},k) & = & \frac{1}{k}e^{ikr_>}\sin
 	 kr_<\nonumber\\
  	& = & \frac{1}{\pi i}\int_{-\infty}^{\infty}\!\! dq\, \frac{e^{iqr}\sin
  	qr^{\prime} }
  	{(q^2-k^2-i\epsilon)}\ .
  	\la{2.3}
\eea

The properties of the Greens function guarantee that $U^{(+)}(r)$
satisfies the Schr\"odinger equation.  The asymptotic form of the
scattering wave at large $r$,
\be
  	\lim_{ r\rightarrow\infty} U^{(+)}(r) =
  	\frac{i}{2}(e^{-ikr}-e^{2i\delta(k)}e^{ikr})
  	\la{2.5}
\ee
and the behavior of the Greens function at large $r$ enable us to read
off the scattering amplitude in terms of $U$,
\be
  	f(k)\equiv e^{i\delta(k)}\sin \delta(k) = \frac{4\pi\lambda}{k}
  	\int_0^\infty\!\! dr\, r\sin kr u(r) \int_0^\infty\!\! dr^{\prime}\,
  	r^{\prime} U^{(+)}(r^{\prime} )u(r^{\prime} )\ .
  	\la{2.6}
\ee

To proceed we return to \eq{2.4}, multiply by $ru(r)$ and integrate
over $r$.  The resulting algebraic equation can be solved for
$\int_0^\infty dr\, r u(r) U^{(+)}(r)$, which yields $f(k)$ upon
substitution into \eq{2.6},
\be
  	f(k) = \frac{4\pi\lambda}{k}\frac{|\xi_0(k)|^2}
  	{1-X(\omega)}
  \la{2.9}
\ee
where we have defined
\bea
  	\xi_0(k)& \equiv & \int_0^\infty\!\! dr\, r u(r) U_0(r) = 
  \int_0^\infty\!\! dr\, r
  	\sin kr u(r) \nonumber\\ X(\omega) & = & 8\lambda\int_0^\infty
  	\!\!dq\,\frac{1}{q^2-\omega-i\epsilon}|\xi_0(q)|^2\ .
  	\la{2.8}
\eea

The scattering amplitude, $f(k)$, has a cut along the positive
$k^{2}$ axis induced by the cut in $X(\omega)$.  The cut begins with a
branch point at threshold, $k^{2}=0$.  The discontinuity across the
cut is given by
\be
 	\mathop{\mathrm disc} f(k) = 2i \Im f(k) = 2i\sin^2\delta(k)
 	\la{2.81}
\ee
where we have used $1/(x+i\epsilon) = {\rm PV} (1/x) -i\pi\delta(x)$ 
to separate out the imaginary part when $x$ is real.  The phase shift 
$\delta(k)$ can be read off \eq{2.9} conveniently by using the 
parameterization, $f(k) = 1/(\cot\delta -i)$,
\be
    \tan\delta(k) =\frac{4\pi\lambda|\xi_0(k)|^2/k}{
    1-8\lambda \pvalt_0^\infty\!\!
    dq\,|\xi_0(q)|^2/(q^2-k^2)}\ .
    \la{2.12}
 \ee
The Born approximation to the scattering amplitude is obtained by 
expanding the denominator in \eq{2.9} in a geometric series,
\be
    f_{\rm BA}(k) =
	\frac{4\pi\lambda|\xi_0(k)|^2}{k}\sum_{n=0}^\infty
    \left[ 8\lambda\int_0^\infty\!\! dq\,\frac{|\xi_0(q)|^2}{q^2-k^2
	-i\epsilon}
    \right]^n\ .
    \la{2.13}
\ee

In addition to the branch cut for real, positive $k^{2}$, $f(k)$ can 
be singular at values of $k$ where the denominator in \eq{2.9} 
vanishes, i.e.,  where $X(\omega)=1$.  Bound states appear as poles in 
$f(k)$ for $k^2<0$, or more precisely, $k=i\kappa$.  [Poles at 
$k=-i\kappa$ are ``virtual states''.]  When $k^2<0$, $X(-\kappa^{2})$ 
becomes real,
\be
    X(-\kappa^{2}) = 8\lambda\int_0^\infty\!\! dq\,
	\frac{|\xi_0(q)|^2}{q^2+\kappa^2}
    \la{2.14}
\ee
and a bound state occurs at the value of $\kappa\equiv\kappa_0$ for 
which
\be
    X(-\kappa_0^{2}) = 8\lambda\int_0^\infty\!\! dq\,
    \frac{|\xi_0(q)|^2}{q^2+\kappa_0^2} = 1\ .
    \la{2.15}
\ee
Since $X(-\kappa^{2})$ is a decreasing function of $\kappa$, the 
criterion for existence of a bound state is that $X(0)>1$, or
  \be
    8\lambda\int_0^\infty\!\! dq\,
    \frac{|\xi_0(q)|^2}{q^2} > 1\ .
    \la{2.16}
  \ee
This equation defines the critical value of $\lambda$ at which a bound 
state appears for a given choice of $u(r)$.  Note that there is at 
most one bound state in this separable potential.  This completes our 
review of the properties of a particle moving in a separable s-wave 
potential.

%%%%%%%%%%%%%%%%%%%%%%%%%%%%%%%%%%%%%%%%%%%%%%%%%%%%%%%%%%%
\section{The Casimir Energy}\label{CasimirEnergy}

We are interested in (a)~the Greens function and (b)~the phase shift
expressions for the Casimir energy.  Both representations may be
derived heuristically in the effective action formalism in field
theory.\cite{fghj}  Equally well, we can start from the formal
expression, \eq{0.1}, and convert the sum over eigenenergies to a
trace of a Greens function or an integral over phase shifts.

\subsection{Greens Function Representation}

The Greens function method starts from the Greens function in
coordinate/energy representation,
\be
	G(\svec r,\svec r^{\prime},\omega)=\sum_j\frac{\phi_j(\svec 
	r)\phi_j^{*}(\svec r^{\prime})}{\omega_j-\omega}
	\la{3.1}
\ee
where the $\{\phi_{j}(\svec r)\}$ are the unit-normalized\footnote{We 
proceed as if the spectrum is discrete and the states are 
normalizable.  Our result applies equally to the case (of interest) 
where the spectrum is continuous and the states are normalized to 
$\delta$-functions.} energy eigenstates in coordinate space and the 
summation ranges over the spectrum of ${\cal H}$, so
\be
	({\cal H}-\omega)G(\svec r,\svec r^{\prime},\omega)=
	\delta^3(\svec r-\svec r^{\prime})\ .
	\la{3.2}
\ee
In order to relate this to the Casimir energy, consider the difference 
of the Greens functions evaluated at $\omega-i\epsilon$ and 
$\omega+i\epsilon$ integrated over space,
\be
	\int_0^\infty\!\! d^3r\, [\Delta G(\svec r,\svec
	r,\omega+i\epsilon)-\Delta G(\svec r,\svec
	r,\omega-i\epsilon)]=2\pi i\sum_j[\delta(\omega-\omega_j)
	-\delta(\omega-\omega_{0j})]
	\la{3.3}
\ee
where $\Delta G$ is the difference between the Greens function in the 
background potential and the free Greens function.  To obtain the 
Casimir energy, multiply \eq {3.3} by $\omega$, integrate from 
$-\infty$ to $\infty$, and divide by $4\pi i$,
\be
	{\cal E}_{c}=
	\frac{1}{4\pi i}\int_{-\infty}^\infty\!\! d\omega\, \omega\int_0^\infty 
	\!\! d^3r\, [\Delta G(\svec r,\svec r^{\prime},\omega+i\epsilon)-\Delta 
	G(\svec r,\svec r^{\prime},\omega-i\epsilon)]\ .
	\la{3.5}
\ee

This result can be simplified further by introducing a function, 
$F(\omega)$, defined by its derivative,
\be
	\frac{dF(\omega)}{d\omega} = \int d^3 r \,\Delta G(\svec r, \svec r,
\omega)
	\la{3.6}
\ee
and the condition that $F(\omega)\to 0$ as $|\omega|\to\infty$.  Then 
substituting into \eq{3.5} and integrating by parts, we find
\be
	{\cal E}_{c}= \frac{1}{4 \pi i}\int_{-\infty}^\infty\!\!\! d\omega\,
	[F(\omega-i\epsilon)-F(\omega+i\epsilon)]\ .
	\la{3.7}
\ee
The surface terms at $\pm\infty\pm i\epsilon$ can be shown to cancel.
		
\begin{figure}
$$
\BoxedEPSF{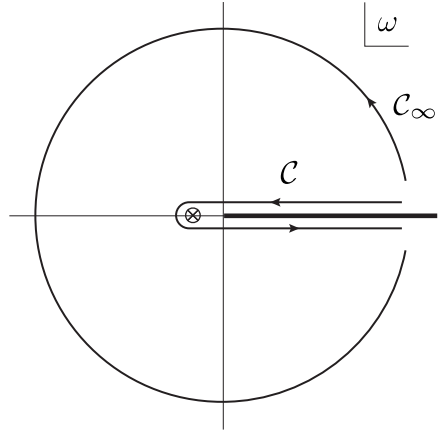} 
$$
\caption{Contours in the complex $\omega$-plane relevant to evaluating
the Casimir energy.}
\label{RJLW:fig:1}
\end{figure}

Referring back to the definition of $\Delta G$, we see that the 
discontinuity in $F$ along the Re $\omega$ axis is associated with the 
eigenstates of ${\cal H}$.  Let $\omega_{0}$ be the ground state of 
${\cal H}$.  For all $\omega<\omega_{0}$ we can send $\epsilon\to 0$.  
Then the integral of \eq{3.7} can be written as a contour integral  over a
contour ${\cal C}$ (as shown in Fig.~\ref{RJLW:fig:1}) that runs in from 
$+\infty$ above the Re $\omega$ axis to $\omega_{0}$ and then returns 
to $+\infty$ below the Re $\omega$ axis,
\be
	{\cal E}_{c}=\frac{1}{4 \pi i} \oint_{\cal C} d\omega\, F(\omega)
	=\frac{1}{4 \pi i} \oint_{{\cal C}_{\infty}}\!\!\! d\omega\, F(\omega)\ .
	\la{3.10}
\ee
In the second term we have replaced the contour ${\cal C}$ by the 
counterclockwise circle at infinity, ${\cal C}_{\infty}$, since in 
general the only singularities in $F(\omega)$ lie on the real axis.  
This compact expression will be particularly useful in the separable 
case.

\subsection{Phase Shift Representation}

The phase shift representation for ${\cal E}_{c}$ begins from \eq{0.1}
divided up into bound state and continuum contributions,
\bea
	{\cal E}_{c}&=&\half\sum_{{\hbox{\scriptsize 
	bound}\atop\hbox{\scriptsize states}} }
	\omega_j + \half\int_{\mbox{\scriptsize continuum}}
\!\!\!\!\!\!\!\!\!\!\!\!\!\!\!	dn\, \omega\nonumber\\
	&=&-\half\sum_jB_j +
\half\int_{0}^{\infty}d\omega\, \rho(\omega)\omega
\la{3.4}
\eea
where $B_j$ are the binding energies, and $\rho(\omega)\equiv 
dn/d\omega$ is the \emph{modification} of the continuum density of 
states due to the interaction.  A simple and general argument connects 
$\rho(\omega)$ to the phase shift (in our case only the s-wave 
contributes),
\be
	\rho(\omega)=\frac{1}{\pi}\frac{d\delta(\omega)}{d\omega}\ .
	\la{3.11}
\ee
Substituting $\rho(\omega)$, and integrating by parts yields,
\be
	{\cal E}_{c}=-\half\sum_jB_j-\frac{1}{2\pi}\int_{0}
	^{\infty} \!\!\! d\omega\,\delta(\omega)\ .
  	\la{3.12}
\ee
The surface term in the integration by parts $\sim 
\omega\delta(\omega)$ vanishes trivially at $\omega=0$.  For a general 
problem the surface term at infinity causes problems because 
generically $\delta(\omega)\sim 1/\omega$ as $\omega\to\infty$.  
However, in the case of a separable potential, we shall see that the 
phase shift vanishes rapidly as $\omega\to\infty$, so the surface term 
can be ignored.  So we may immediately calculate the Casimir energy if 
we are given the phase shift.  \Eq{3.10} and \eq{3.12} give 
alternative representations of the Casimir energy, which should both be 
equivalent to \eq{0.5} for the separable case.

%%%%%%%%%%%%%%%%%%%%%%%%%%%%%%%%%%%%%%%%%%%%%%%%%%%%%%%%%%%%%%%%%%%%%%

\section{Casimir Energy for the Separable
Case}\label{CasimirEnergySeparable}

In this section we examine the Green's function and phase shift 
representations of the Casimir energy for the separable case.

\subsection{Greens Function Representation}

To exploit \eq{3.5} we need an explicit expression for the trace of 
the Greens function for the separable potential of Section 2.  We 
start with the Lippmann-Schwinger equation for $G$,
\be
  	G(\svec r, \svec r^{\prime},\omega) = G_0(\svec r, \svec r^{\prime},\omega)
  	+ \lambda \int d^{3}r_{1}\, d^{3}r_{2}\, G_{0}(\svec r, \svec r_{1}, 
  	\omega) u(r_{1})u(r_{2})G(\svec r_{2},\svec r',\omega)
  	\la{4.0}
\ee
which may be solved by iteration,
\be
  	\Delta G(\svec r, \svec r^{\prime},\omega)
  	= \lambda\int d^3r_1\, d^3r_2 \,G_0(\svec r, \svec r_1,\omega)
  	u(\svec r)\left[ 1\! +\! X(\omega)\! +\!
  	X(\omega)^2\!+\cdots\right]u(\svec r^{\prime})
  	G_0(\svec r_2,\svec r^{\prime},\omega)
  	\la{4.4}
\ee
where $X(\omega)$ was defined in \eq{2.8}.

\Eq{3.5} requires the us to set $\svec r = \svec r'$ and integrate.  
If we substitute from \eq{4.4} the result takes a particularly simple 
form,
\bea
 	 \int d^3r \,\Delta G(\svec r, \svec r, \omega) &=
	&\frac{dX}{d\omega}\sum_{m=0}^\infty
  	\frac{[X(\omega)]^m}{m}\nonumber\\
  	&=& -\frac{d}{d\omega} \ln (1 - X(\omega))
  	\la{4.6}
\eea
so
\be
  	{\cal E}_{C}=-\frac{1}{4\pi i}\oint_{{\cal C}_{\infty}}
  	d\omega\, \ln (1 - X(\omega))
  	\la{4.7}
\ee
where we have identified the generic function 
$F(\omega)$ of \eq{3.6} with $-\ln(1-X(\omega))$ in the separable 
case.  [With this identification it is easy to check that the surface 
terms discussed following \eq{3.7} do indeed vanish.]  To evaluate 
the integral we need $\lim_{\omega\to\infty}\ln(1-X(\omega))$.  From 
\eq{2.8} and the normalization of $u(r)$, it is easy to see that as 
$\omega\to\infty$,
\be
  \ln(1-X(\omega)) \sim -X(\omega) \sim \lambda/\omega\ .
  \la{4.7a}
\ee
Thus,
\bea
{\cal E}_{c}=-\frac{\lambda}{4 \pi i}\oint_{C_\infty}
\frac{d\omega}{\omega}=-\frac{\lambda}{2}\ .
  \la{4.8}
\eea
Confirming the value of the Casimir energy is $-\lambda/2$ as we 
derived in Section 1 from more formal considerations.

Some comments are in order:
\bit
\item The method we have presented is equivalent to the usual, 
graphical analysis of the effective action in quantum field theory.  
Had we begun with a theory describing a scalar field, $\Psi$, 
propagating in a scalar background, $\Phi$, then the Casimir energy 
would have been given by the (infinite) sum of one-loop graphs for 
$\Psi$ with insertions of the $\Psi/\Phi$ coupling.  The one-loop 
graphs could then be resummed into a Lippmann Schwinger equation for 
$G(\svec r, \svec r', \omega)$.  The special advantages of a separable 
potential are (a)~that the resulting integral equation for $G$ is 
solvable, and (b)~that there are no ultraviolet divergences (seen as 
divergences in the $\omega\to\infty$ limit) to complicate the 
calculation.

\item The resummation of $1+X+X^{2}+\cdots$ that generated the 
logarithm in \eq{4.6} is valid for small $\lambda$ where the series 
converges.  The result can then be analytically continued to large 
$\lambda$ were the re-expansion into a (Born) series does not converge.

\eit

%%%%%%%%%%%%%%%%%%%%%%%%%%%%%%%%%%%%%%%%%%%%%%%%%%%%%%%%%%%%%%
\subsection{Phase Shift Representation}

In Section 3 we also derived a simple representation, \eq{3.12}, for 
the Casimir energy as a sum over binding energies and an integral over 
the phase shift $\delta(\omega)$.  In this subsection we show 
explicitly that this representation is equivalent to the Green's 
function representation as found, for example, in \eq{4.7}.

First, we return to the contour, ${\cal C}$, of Fig.~\ref{RJLW:fig:1},
\be
  	{\cal E}_{c}=-\frac{1}{4\pi i}\oint_{\cal C}
  	d\omega\, \ln (1 - X(\omega)).
  	\la{4.9}
\ee
The contour integral is equivalent to integrating the discontinuity in 
the integrand across the real axis from just below the lowest bound 
state $\omega_{0}$ to $\infty$,
\be
  	{\cal E}_{c}=\frac{1}{2\pi}\int_{\omega_{0}}^{\infty}
 \!\!\!	d\omega\, {\rm Im}\ln (1 - X(\omega+i\epsilon))
  	\la{4.10}
\ee
where we have used the fact that the discontinuity in the logarithm is 
$2i$ times its imaginary part.

There are two distinct regions in the integral.  For $\omega<0$, 
$X(\omega)$ is explicitly real, and the logarithm can have an 
imaginary part if and only if $X(\omega)>1$.  For $\omega>0$, 
$X(\omega)$ always has an imaginary part (see \eq{2.8}) related to the 
scattering amplitude.  We shall show that these two contributions map 
into the binding energy and integral over the phase shift respectively, 
as expected from \eq{3.12}.

First consider $\omega\le 0$.  According to the analysis of Section 2, 
\eqs{2.14}{2.15}, etc., $X(\omega)$ takes its maximum at $\omega = 0$, 
and decreases as $\omega$ decreases.  If $X(0)>1$ there is a single 
bound state with binding energy, $B\equiv-\omega_{0}$, determined by 
the equation $X(-B)=1$.  Thus if $\lambda$ is large enough to generate 
a bound state, then $\ln(1-X(\omega))$ has an imaginary part (equal to 
$\pi$) for $-B\le \omega \le 0$.  Thus
\be 
	\frac{1}{2\pi}\int_{\omega_{0}}^{0}d\omega\, {\rm Im}\ln (1 - 
	X(\omega)) = -\half B\ .
	\la{4.11}
\ee

Next consider $\omega>0$.  From the analysis of Section 2, we find
\be
	\ln(1-X(\omega))=\ln\left(1-4\lambda
	\pval_{-\infty}^{\infty}dq\,
	\frac{|\xi_0(q)|^2}{(q^2-\omega)}+ 4\lambda i \pi
	\frac{|\xi_0(k)|^2}{k}\right)\ .
	\la{4.14}
\ee
From this we read off the imaginary part,
\be
    {\rm Im} \ln(1-X(\omega+i\epsilon))=-\tan^{-1}
	\left(\frac{4\pi\lambda|\xi_0(k)|^2/k}{
    1-4\lambda\pvalt_0^\infty
    dq\,|\xi_0(q)|^2/(q^2-\omega)}\right)
    \la{4.15}
 \ee
which is just $-\tan\delta(k)$ as defined in \eq{2.12}.  Substituting 
from \eqs{4.11}{4.15} into \eq{4.10}, we confirm the phase shift plus 
binding energy representation, \eq{3.12},
\be
	{\cal E}_{c}=-\half\sum_jB_j-\frac{1}{2\pi}\int_{0}
	^{\infty} \!\!\! d\omega\,\delta(\omega)\ .
  	\la{4.16}
\ee
%
%%%%%%%%%%%%%%%%%%%%%%%%%%%%%%%%%%%%%%%%%%%%%%%%%%%%%%%%%%%%%%
\section{Convergence and Numerical Issues in a Separable Potential 
Model}\label{Convergence}

A particular virtue of the separable potential model is that it is 
simple enough to allow us to investigate issues that are difficult to 
attack in more realistic theories.  As an example of such an issue, we 
consider here some computational aspects of phase shift representation 
of the Casimir energy.  Ref.~\cite{fghj} introduced a method of 
computing the Casimir energy in realistic quantum field theories based 
on the Born approximation to the phase shift.  Divergences generated 
by the lowest-order Feynman diagrams are associated with the first few 
terms in the Born approximation to the phase shift.  To remove the 
divergences from the numerical part of the calculation, the first few 
Born approximations are subtracted from the phase shift.  These 
contributions are then added back into the calculation in the form of 
(divergent) Feynman diagrams, which are regulated and renormalized by 
means of available counterterms.  Schematically
\be
   {\cal E}_{c} = 
   -\half\sum_{j}B_{j}-\frac{1}{2\pi}\int_{0}^{\infty}\!\!\! d\omega\,
   \bar\delta^{N}(\omega) +\sum_{n=0}^{N} D_{n} + {\rm CT}
   \la{5.1}
\ee
where $\bar\delta^{N}$ is the phase shift with the first N terms in the 
Born approximation subtracted,
\be
	\bar\delta(\omega) = \delta(\omega) 
	-\sum_{n=1}^{N}\lambda^{n}\delta^{(n)}(\omega)
	\la{5.2}
\ee
$D_{n}$ is the contribution to ${\cal E}_{c}$ from the $n^{\rm th}$-order
Feynman diagram, and CT denotes the renormalization  counterterms. 
Both the $\{D_{n}\}$ and the counterterms depend on a  cutoff if the
theory has ultraviolet divergences.  The  Born-subtracted phase shift,
however, is cutoff independent.

In realistic quantum field theory applications the Feynman diagrams, 
$\{D_{n}\}$, and the counterterms are calculated by standard, 
diagrammatic methods.  The integral over $\bar\delta^{N}$ is done 
numerically.  The Born approximations $\{\lambda^{n}\delta^{(n)}\}$ 
have just the right behavior to cancel the large $\omega$ tail of 
$\delta(\omega)$ and render the $\omega$-integral convergent.  However 
they are a poor approximation to $\delta(\omega)$ for small $\omega$, 
especially in partial waves where there are bound
\hbox{states.\cite{fgjw}}  Numerical studies show that at low $\omega$
the function 
$\bar\delta^{N}(\omega)$ is much larger than its integral 
$\int_{0}^{\infty}d\omega\, \bar\delta^{N}(\omega)$.  The origin of this 
effect is obscured in those studies because the integral over the 
individual Born approximations, $\int_{0}^{\infty}d\omega\, 
\delta^{(n)}(\omega)$, diverge.  The large $\omega$ behavior of the 
separable potential model is very soft, enabling us to study this 
issue explicitly.

It is convenient to specialize further to a particular choice of the 
separable potential function, $u(r)$.  Following Gottfried, we choose 
the Yukawa function,
\be
	u(r) = \sqrt{\frac{\alpha}{2\pi}}\frac{e^{-\alpha r}}{r}\ .
	\la{5.3}
\ee
A simple calculation gives
\bea
	\xi_{0}(k)&=&\sqrt{\frac{\alpha}{2\pi}}
	\frac{k}{k^{2}+\alpha^{2}}\nonumber\\
	X(k)&=& \frac{\lambda}{2(\alpha -ik)^{2}}\ .
	\la{5.4}
\eea
It is convenient to use the momentum, $k$, rather than $\omega=k^{2}$ as 
the independent variable, also to introduce scaled variables, 
$k/\alpha\equiv z$, $\lambda/\alpha^{2}\equiv g$, and finally, it is 
easiest to explore the model by studying the scattering amplitude, 
$f(k)\to f(g,z) = 1/(\cot\delta(g,z) - i)$,
\be
	f(g,z) = \frac{2gz}{(1+iz)^{2}((1-iz)^{2}-g)}\ .
	\la{5.5}
\ee

\begin{figure}
$$
\BoxedEPSF{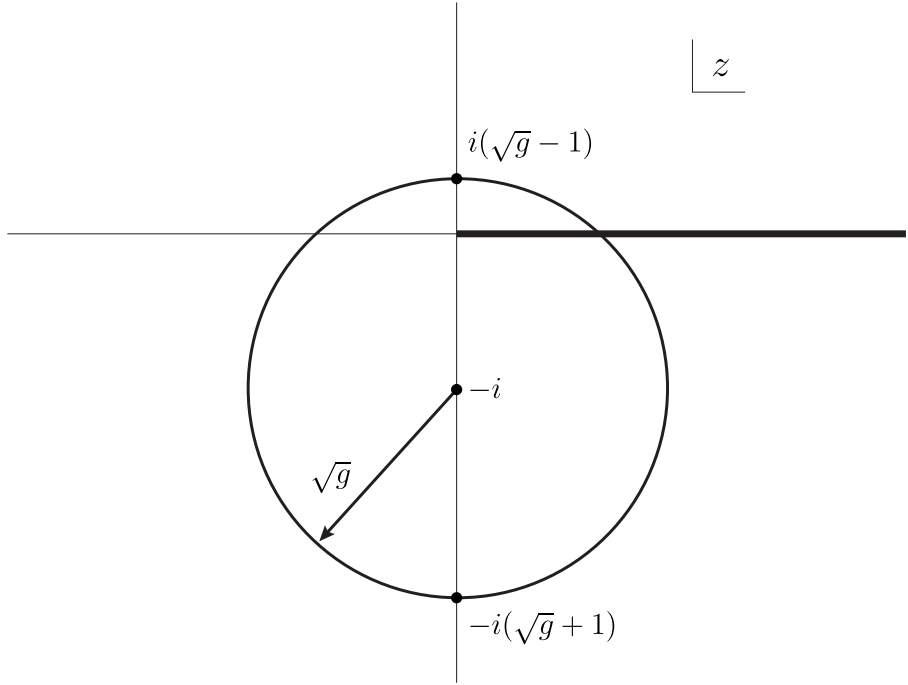 scaled 1500} 
$$
\caption{The Born expansion converges outside a circle of radius $\sqrt
g$ centered at $-i$ in the complex $z$-plane. Singularities in $f(g,z)$ at
$z=-i$, $z_+ = i(\sqrt g -1)$, and $z_- = -i(\sqrt g +1)$ are shown.}
\label{RJLW:fig:2}
\end{figure}

The scattering amplitude $f(g,z)$ has two types of singularities.  The 
double pole at $z=-i$ is a ``potential singularity'' arising from the 
fourier transform of $u(r)$, $\xi_{0}(k)$.  It is on the second sheet 
of the complex $\omega$ plane where it does not affect the contour 
shifting arguments we used earlier in this paper.  The other 
singularities are poles at
\be
	z_{\pm}=\left\{  \begin{array}{l}
	i(\sqrt{g}-1)\\
	-i(\sqrt{g}+1).
	\end{array}
	\right. \ .
	\la{5.6}
\ee
For $g>1$, $z_{+}$ is associated with a bound state: it lies on the 
positive imaginary axis, corresponding to the negative real axis on 
the first sheet of the complex $\omega$-plane.\footnote{For $g< 1$ the 
$z_{+}$ pole lies on the negative imaginary axis, i.e.,  on the second 
sheet of the $\omega$-plane.  As $g\to 1$ it becomes virtual state on 
its way to becoming a bound state.} The energy of the bound state is 
$\omega_{0} = -\alpha^{2}z_{-}^{2}= - (\sqrt{\lambda}-\alpha)^{2}$.  
The pole at $z_{-}$ is always on the second sheet and is not 
dynamically important.  The singularities in the complex-$z$ plane are 
summarized (for $g>1$) in Fig.~\ref{RJLW:fig:2}.

With these considerations in mind, we can write an illuminating 
expression for the Casimir energy,
\be
	{\cal E}_{c} = -\half\lambda = 
	\left\{  \begin{array}{ll}\displaystyle
	-\frac{1}{2\pi}\int_{0}^{\infty} \!\! d\omega\, \delta(\omega) & 
	\quad \mbox{for $\lambda <\alpha^{2}$}\\[1ex]
\displaystyle
	-\half(\sqrt{\lambda}-\alpha)^{2} 
	-\frac{1}{2\pi}\int_{0}^{\infty}\!\! d\omega\, \delta(\omega) &
	\quad \mbox{for $\lambda >\alpha^{2}$}
	\end{array}
	\right.
	\la{5.7}
\ee
i.e.,  where there is a bound state, it appears explicitly in ${\cal 
E}_{c}$.  The Born approximation is an expansion of $\delta(g,z)$ in 
powers of $g$ obtained by expanding the denominator of $f(g,z)$ (see 
\eq{5.5}) in a geometric series:
\be
	\delta(g,z) = \sum_{n=1}^{\infty} g^{n}\delta^{(n)}(z)\ .
	\la{5.8}
\ee
The convergence of the Born expansion is determined by the convergence 
of the geometric series for $f(g,z)$, which converges provided 
$|1-iz|>\sqrt{g}$.  That is, $z$ must lie outside the circle of radius 
$\sqrt{g}$ centered at $z=-i$.   To compute 
the Casimir energy by the phase shift method, we must integrate over 
all $\omega>0$, corresponding to $z\in [0,\infty]$.

\begin{figure}
$$
\BoxedEPSF{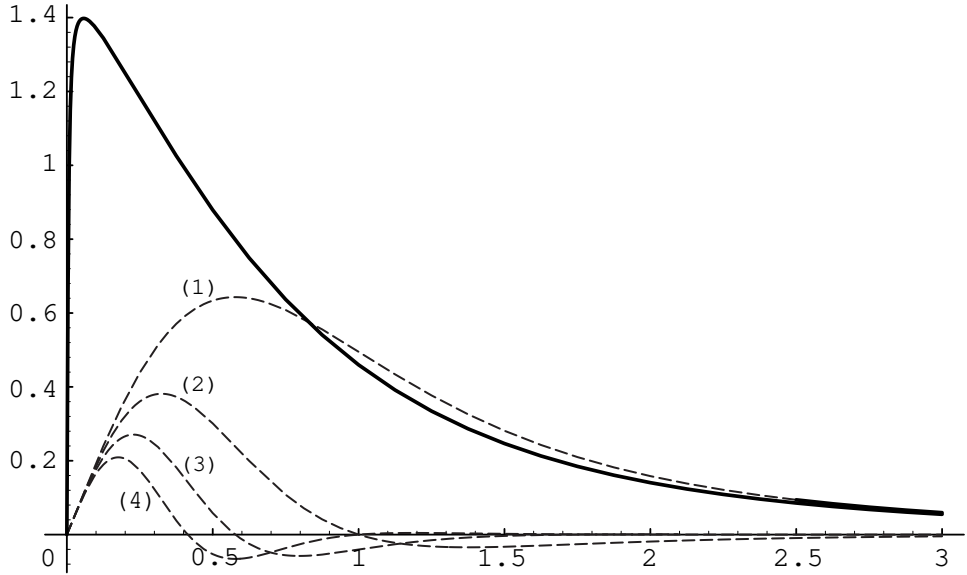 scaled 750} 
$$
\caption{The phase shift for the Yukawa separable potential for
$g= 0.99$.  The solid curve is the exact result, the four dashed
curves give the first four Born approximations respectively.}
\label{RJLW:fig:3}
\end{figure}

As long as $g<1$ ($\lambda < \alpha^{2}$) the Born expansion converges 
for all real, positive $z$ and can be used under the integral sign in 
\eq{5.7} to compute ${\cal E}_{c}$,
\be
	{\cal E}_{c} = -\half \lambda = -\frac{1}{2\pi}\sum_{n=1}^{\infty}
	\lambda^{n}\int_{0}^{\infty}\!\! d\omega\, \delta^{(n)}(\omega)
	\qquad \mbox{for}\ \lambda < \alpha^{2}\ .
	\la{5.9}
\ee
This is a remarkable result: both sides are power series in $\lambda$, 
but the left hand side consists of a single term.  We conclude that, 
for $\lambda<\alpha^{2}$, when there is no bound state, the entire 
Casimir energy is given by the first Born approximation, and that the 
integrals over all higher Born approximations to the phase shift 
vanish:
\bea
	{\cal E}_{c} &=& -\half \lambda = -\frac{\lambda}{2\pi}
	\int_{0}^{\infty}\!\! d\omega\, \delta^{(1)}(\omega),\nonumber\\
	0&=& \int_{0}^{\infty}\!\! d\omega\, \delta^{(n)}(\omega)
	\qquad \mbox{for all}\ n>1\ .
	\la{5.10}
\eea
This result may easily be verified numerically.  

When $g>1$ ($\lambda > \alpha^{2}$), it is no longer possible to use 
the Born approximation to evaluate the $\omega$-integral.  A glance at 
the second equation in \eq{5.7} confirms that the Born approximation 
cannot converge when there is a bound state: if it did, it would yield 
a result analytic in $\lambda$, however the $\omega$-integral must 
generate a factor $\alpha\sqrt{\lambda}$ to cancel the contribution of 
the bound state and give ${\cal E}_{c}=-\half \lambda$.  To generate 
$\sqrt{\lambda}$, the Born expansion must not converge.  Still, it is 
useful to subtract some number of terms in the Born expansion from the 
exact phase shift ($\bar\delta^{N}(\omega)=\delta(\omega)- 
\sum_{n=1}^{\infty}\lambda^{n}\delta^{(n)}(\omega)$) and to rewrite ${\cal 
E}_{c}$ accordingly,
\bea
	{\cal E}_{c} = -\half\lambda &=&
	-\half(\sqrt{\lambda}-\alpha)^{2} -\frac{\lambda}{2\pi}\int_{0}^{\infty}
\!\!	d\omega\, \delta^{(1)}(\omega) -\frac{1}{2\pi}\int_{0}^{\infty}
\!\!	d\omega\, \bar\delta^{N}(\omega)\nonumber\\
	&=& -\half(\sqrt{\lambda}-\alpha)^{2} - \half\lambda 
	-\frac{1}{2\pi}\int_{0}^{\infty}
\!\!	d\omega\, \bar\delta^{N}(\omega)\ .
	\la{5.11}
\eea
Here we have subtracted the first $N$ terms in the Born 
approximation to convert $\delta$ to $\bar\delta^{N}$.  When we added 
them back in, we used the fact that the integral of the first Born 
approximation gives exactly $-\half\lambda$, and that the integrals of 
all higher Born approximations vanish.  From this analysis we conclude 
that when there is a bound state, the integral of the ``Born 
subtracted'' phase shift, $\bar\delta^{N}$ gives an $N$-independent 
contribution to ${\cal E}_{c}$,
\be
	-\frac{1}{2\pi}\int_{0}^{\infty}\!\!
	d\omega\, \bar\delta^{N}(\omega) =
	\half(\sqrt{\lambda}-\alpha)^{2}\ .
	\la{5.12}
\ee

\begin{figure}
$$
\BoxedEPSF{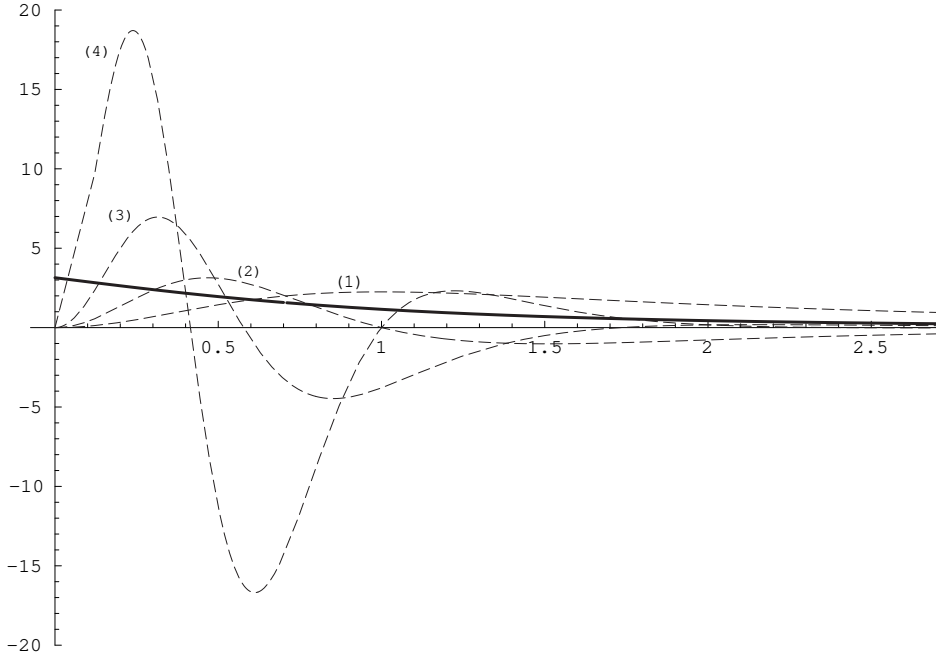 scaled 750} 
$$
\caption{The phase shift for the Yukawa separable potential for
$g= 4.5$.  The solid curve is the exact result, the four dashed
curves give the first four Born approximations respectively.}
\label{RJLW:fig:4}
\end{figure}

These results help to understand the effects that have been seen in 
numerical calculations in more complicated theories.  In our toy  model,
when $g$ is such that there is no bound state, the Casimir  energy is
given entirely by the first Born approximation (which, in a  theory with
divergences, would be computed from a Feynman graph).  The  
$n^{\rm th}$
Born approximation gives a contribution to $\delta$ of order 
$g^{n}$, which may be large for $g\lesssim 1$, but its integrated 
contribution to the Casimir energy is exactly zero.  In more  complicated
theories the answer is very small but not exactly  zero.\cite{fgjw} The
magnitude of the effect can be seen in Fig.~\ref{RJLW:fig:3},  where we
plot the exact phase shift and the first four Born  approximations for
$g=0.99$.  The first Born approximation has the same integral as the full
phase shift. Each higher Born approximation integrates to zero. The
situation becomes even more bizarre  in when there is a bound state. 
The entire Casimir energy is still  given by the first Born approximation! 
The bound state contribution  is exactly canceled by the remainder of
the phase shift integral.   One can remove any number of further Born
approximations from $\delta$  -- defining
$\bar\delta^{N}$, which is significantly modified by the  subtraction of
higher Born approximations -- without changing the  integral.  The effect
can be seen in Fig.~\ref{RJLW:fig:4}, where we plot the exact  phase shift
and the first four Born approximations for $g= 4.5$.  Once again the toy
model caricatures effects seen in more  complex and more realistic
theories.

%%%%%%%%%%%%%%%%%%%%%%%%%%%%%%%%%%%%%%%%%%%%%%%%%%%%%%%%%%%%%%
\section{Summary and Conclusions}\label{Summary}

Our principal purpose has been to develop a simple model to obtain 
insight into complicated calculations of Casimir energies in quantum 
field theories.  We have verified in our separable potential model 
that several traditional representations of the Casimir energy 
employed in quantum field theories are equivalent.  Our model is very 
simple.  In particular, it does not display the divergences 
characteristic of real quantum field theories.  Thus we are not 
surprised that results are confirmed in our model that might be 
spoiled by divergences in more realistic theories.  Nevertheless, the 
model is rich enough to illustrate some of the computational 
difficulties that have been seen numerically in more complicated 
theories.

Also, we should point out that there are cases where separable 
potentials do arise in the treatment of more realistic theories.  In 
particular, Bashinsky has shown that the collective coordinate 
quantization in theories with solitons introduces additional, 
separable-potential-like terms into the small oscillations Hamiltonian 
whose eigenvalues determine the Casimir energy.\cite{Bashinsky} The 
ideas developed here would apply relatively directly to such cases.

%%%%%%%%%%%%%%%%%%%%%%%%%%%%%%%%%%%%%%%%%%%%%%%%%%%%%%%%%%%%%%
\section*{Acknowledgments}

We thank S.~Bashinsky, J.~Goldstone, N.~Graham, and H.~Weigel for 
conversations and suggestions related to this work.

%%%%%%%%%%%%%%%%%%%%%%%%%%%%%%%%%%%%%%%%

%%%%%%%%%%%%%%%%%%%%%%%%%%%%%%%%%%%%%%%%%%%%%

\end{document}